\documentstyle[aps,prl,twocolumn,epsf]{revtex} 

\begin{document} 

\draft

\title{Negative length orbits in normal-superconductor
billiard systems}

\author{J. Cserti, G. Vattay, J. Koltai\\ 
E\"otv\"os University, Department of Physics of 
Complex Systems, \\
H-1117 Budapest, P\'azm\'any P\'eter s\'et\'any 1/A, Hungary\\
F. Taddei and C. J. Lambert\\
School of Physics and Chemistry, Lancaster University\\
Lancaster LA14YB, United Kingdom}

\wideabs{

\maketitle

\begin{abstract}
The Path-Length Spectra of mesoscopic systems including diffractive 
scatterers and connected to superconductor is studied theoretically. 
We show that the spectra differs fundamentally from that of normal systems 
due to the presence of Andreev reflection.   
It is shown that negative path-lengths should arise in the spectra 
as opposed to normal system.
To highlight this effect we carried out both quantum mechanical 
and semiclassical calculations for the simplest possible diffractive scatterer.
The most pronounced peaks in the Path-Length Spectra of the reflection 
amplitude are identified by the routes 
that the electron and/or hole travels.
\end{abstract}

}

In recent years, semiclassical methods have become a popular tool for
describing devices operating in the mesoscopic regime. Advances in 
manufacturing and material design have made possible the creation of clean 
mesoscopic devices, whose properties depend on the microscopic details 
of individual samples. For example,
in recent experiments\cite{6,7} involving 
semiconductor microjunctions, both the quantum coherence 
length and the mean free path of elastic collisions are large 
compared to the size of the junction.
In such devices electrons can be described as
a two-dimensional ideal Fermi gas of noninteracting particles. 
The conductance of such junctions has been measured and
found to oscillate strongly as the Fermi energy is varied. 

Semiclassical methods have proved to be very effective for 
understanding conductance fluctuations in normal 
microjunctions. On the one hand,
methods based on random matrix theory
\cite{4,5,Carlo1} successfully predict statistical 
properties of transport properties. On the other,
short-wavelength semiclassical descriptions
able to explore geometry-induced interference effects in weakly
disordered or clean mesoscopic devices. 
In particular, it has been shown that the {\em Path-Length Spectra}
(PLS), defined as the 
power spectrum of the reflection (transmission) amplitudes
with respect to the Fermi wavelength  
\begin{equation}
\hat{r}_{mn}(L)={\left| 
\int_{k_{min}}^{k_{max}}e^{-ik_FL}r_{mn}(k_F)dk_F \right| }^2,
\label{PLS}
\end{equation}
possesses peaks at lengths corresponding to classical trajectories
of electrons starting and ending at the external contacts
\cite{B1,B2,Burgerdorf1,Burgerdorf2,Burgerdorf3,Delos,Lin-semi}. 
Here $r_{mn}(k_F)$ ($t_{mn}(k_F)$) is
the reflection (transmission)
amplitude at the Fermi wavenumber $k_F$ for
scattering from mode $n$ of the entrance lead to mode $m$ of the entrance
(exit) lead in a two probe conductance measurement.

Mesoscopic devices connected to a superconductor
present a new challenge for semiclassics. 
In such normal-superconductor (NS) systems Andreev
reflection\cite{Andreev,Carlo1,Carlo-konyv,Colin1} 
plays an important role, whereby
electrons at the Fermi energy in the normal metal are 
retro-reflected as holes at the NS interface.
Such a process might be expected to dramatically affect the PLS, but
to-date, no investigations of the PLS in NS systems have been carried out.

In this Letter we present the first such investigation, by examining
a mesoscopic device connected
to a superconductor and show that the PLS of NS systems differ from
those of normal systems in a fundamental way. In particular
for NS systems containing diffractive scatters,
negative path-lengths can arise in the PLS which are absent from the
corresponding normal systems.

Before turning to the NS system we shortly discuss the role of diffraction
in the PLS of normal systems, following Ref.\cite{Chaos-cikk}.
In the semiclassical approximation, particles hitting a diffractive 
scatterer  may scatter in any direction since the classical 
dynamics is not uniquely defined \cite{Keller,Diffract}.  
In Ref.\cite{Chaos-cikk} a two-dimensional 
wave guide with a small point-like diffractive scatterer has been analyzed
\cite{Exner} 
(see Fig 1a.). 
It has been shown that the PLS of the reflection
amplitude $r_{nm}$ has peaks at path-lengths corresponding 
to classical trajectories starting and returning to the entrance of the
lead, either diffracted once or several times by the scatterer.
At large Fermi wavelengths a typical trajectory with multiple bounces is
shown in Fig.\ 1a. Such trajectories consist of two parts:
segments (along the z axis) connecting the lead
and the scatterer with total length $2 \times z_0$, and  
multiple diffraction trajectories starting and ending on the scatterer.
A multiple diffraction trajectory  can be decomposed into loops that 
start and end on the scatterer, making bounces on the walls of the waveguide.
The possible lengths of the loops are
\begin{equation}
l_r=
\left\{
\begin{array}{c}
2Wr\\
2x_0+2Wr\\
2(W-x_0)+2Wr\\
\end{array}
\right. ,
\label{lr}
\end{equation}  
where $r=0,1,2,...$ is the repetition number and $W$ is the width of 
the waveguide \cite{Chaos-cikk}. 

In Figure \ref{Q-szemi-N} the PLS calculated quantum mechanically
and semiclassically using the method developed in Ref.\cite{Chaos-cikk} 
are shown. 
One can see that the agreement between the quantum and the
semiclassical calculation is excellent and the peaks are located
at lengths which are linear combinations of (\ref{lr}) plus $2z_0$. 
It is obvious that the PLS 
has peaks {\em only} for positive lengths $L$. The amplitude of the peaks 
is decreased by multiple diffraction and therefore the most pronounced 
peaks correspond to paths diffracted only once on the scatterer.   

Now consider the effect of replacing one of the
exit leads by a superconductor.
In this case a new contribution to the reflection amplitude $r_{mn}$ has to be
taken into account, namely the one coming from Andreev reflection at
the NS interface.
By solving the Bogoliubov-de Gennes equation for a NS interface it is possible
to show that, at the Fermi energy, electron-like excitations impinging onto the
superconducting interface are coherently retro-reflected as hole-like
excitations.
In a semiclassical description,
the classical action associated with the path connecting points $q'$ and ${q''}$
is given by
\begin{equation}
S(q',{q''})=\int_{q'}^{{q''}}p(q)dq,
\end{equation} 
where $p(q)$ is the momentum of the electron or the hole along the path (note
that if the path touches the superconductor at least once, it will contain
both electron and hole parts). In particular the action associated with an
Andreev
reflection process in which an electron starting at point $q'$ returns to
the same point as a hole
can be written as
\begin{equation}
S(q',{q'})=k_F L(q',q^*) - k_F L(q^*,{q'}) = 0,
\end{equation} 
where $L(q',q^*)$ denotes the length of the path of the electron
until it hits the superconductor at $q^*$ and $L(q^*,{q'})$ is
the path length of the hole from $q^*$ to ${q'}$. 
The minus sign in the second term is due to the fact that
the directions of the momentum  and the velocity of the hole are
opposite. The total action
of an electron-hole (e-h) trajectory returning to its starting point
is always zero, since the hole retraces the path of the electron.
As a consequence of this result, the PLS of the reflection
amplitude for electron-electron (e-e) has peaks at positive lengths $L$
while for e-h
it has a pronounced peak at $L=0$.   

We now consider the case where {\em diffractive scattering}
is possible in the system (Fig.\ 1b). The new feature is that the trajectory
of electrons or holes hitting a diffractive point is not uniquely defined.
Therefore, a hole retracing an electron, which scattered on a
diffractive center, will not necessarily retrace the 
trajectory of the electron beyond the diffractive center. The hole
may leave the diffraction center at a different angle than that of the
incident electron.
This effect has already been pointed out by Beenakker\cite{BB} 
in connection with normal-metal-superconductor junction
containing a point contact. Consequently, in the presence of diffraction,
complicated trajectories consisting of several electron and hole segments
may arise. The classical action for this case can be written as a sum
of actions of the segments
\begin{equation}
S=\sum k_F (\pm L_i),
\label{S}
\end{equation}
where $L_i$ is the length of the segment $i$ and the $+$ or $-$ 
correspond to cases when electron or hole traverses the segment $i$,
respectively. Unlike in the diffractionless case, the sum
of positive and negative terms in Eq.\ (\ref{S}) is not necessarily zero.
Moreover, the total
length of the hole segments may exceed those of the electrons. Thus, in the PLS
of the e-e reflection amplitude, peaks at {\em negative lengths} may
appear, which are completely absent from the PLS of the corresponding
normal systems. 

To observe negative lengths in the PLS we note that when the exit lead is
replaced by a
superconductor \cite{Carlo1} the e-e reflection amplitude can be
expressed in terms of the transmission and reflection amplitudes
of the corresponding normal system (see Eq.\ 266a in Ref.\cite{Carlo1}).
For the system of Fig.\ 1b,
the latter amplitudes have been derived in 
Ref.\cite{Chaos-cikk} and, after lengthy but straightforward algebra, 
we find the
following expression
for the $n,m$ matrix element of the e-e reflection amplitude at the Fermi
energy:
\begin{equation}
s_{nm}^{e-e}(k_F)=\frac{2 r_{nm}(k_F)}
{1+|{\cal D}|^2 \left[Im \, G_0(x_0,z_0|
x_0,z_0)\right]^2},
\label{hetes}
\end{equation}    
where $r_{nm}(k_F)$ is the matrix of reflection amplitudes of the
normal system\cite{Chaos-cikk} for the entrance lead, $G_0(x,z|x',z')$ is
the Green's function of the empty waveguide. 
${\cal D}=-i\tilde{\lambda}/
\left[1-\tilde{\lambda} G_0(x_0,z_0|x_0,z_0)\right ]$ 
can be regarded as the diffraction constant of the scatterer, 
where $\tilde{\lambda}$ is the
renormalized strength of the scatterer\cite{Crossover}.  

On the upper part of Fig.\ \ref{Q-szemi-S} 
the PLS of the reflection amplitude of (\ref{hetes}) (the exact 
quantum mechanical result) is plotted as a function of the path length. 
Peaks at negative path lengths are clearly visible here as opposed to 
normal system shown in Fig.\ \ref{Q-szemi-N}. 

The semiclassical approximation of Eq.\ (\ref{hetes}) can be obtained 
from the semiclassical form of the Green's function which is given by
\begin{equation}
G_0(x_0,z_0|x_0,z_0) = \sum \frac{{(-1)}^{n_r}}
{\sqrt{8\pi k_F l_r}} \, e^{ik_F l_r -i3\pi /4},
\label{G0-szemi}
\end{equation} 
where the summation is over all possible loops with 
lengths $l_r$ given in (\ref{lr}) and $n_r$ is the number of bounces 
on the walls of the waveguide.
One can see that $Im \, G_0(x_0,z_0|x_0,z_0)$ will contain the 
complex conjugate of Eq.\ (\ref{G0-szemi}) and is therefore a sum with 
terms proportional to $e^{\pm i k_F l_r}$. Here, terms 
with $e^{-i k_F l_r}$ correspond to loops traversed by holes.
The e-e reflection amplitude  given in (\ref{hetes})
can be rewritten as a multiple diffraction series:  
\begin{eqnarray}
s_{nm}^{e-e}(k_F) &=& 2r_{nm}(k_F) 
\left\{ 1- |{\cal D}|^2 \left[Im \, G_0(x_0,z_0|x_0,z_0)\right]^2  
\right.  \nonumber \\ 
 &+& \left. |{\cal D}|^4 \left[Im \, G_0(x_0,z_0|
x_0,z_0)\right]^4 \cdots\right\}. 
\end{eqnarray}
Expressing powers of $Im\, G_0$ with  Eq.\ (\ref{G0-szemi}) and its 
complex conjugate $s_{nm}^{e-e}(k_F)$ can be written as an oscillating sum 
\begin{equation}
s_{nm}^{e-e}(k_F)= \sum_j A_j e^{iS_j},
\label{szemi-s-ee}
\end{equation}
where $S_j$ is of form Eq.\ (\ref{S}). The amplitudes $A_j$ 
can be determined exactly by using the above formulas. 
In the lower part of Fig.\ \ref{Q-szemi-S} the PLS of the reflection 
amplitude $s_{nm}^{e-e}(k_F)$ is plotted using Eq.\ (\ref{szemi-s-ee}). Again
one can see a very good agreement between the quantum and semiclassical 
calculations. The most pronounced peaks with negative lengths 
(see Fig.\ \ref{Q-NS}) 
come from the family $L= 2z_0 -l_r$, where $l_r$ is 
given in Eq.\ (\ref{lr}), namely 
$L=-0.4,-1.0,-1.6,-2.4,-3.0,-3.6,-4.4,-5,\cdots$ .
These lengths can be associated with the following routes:
First the electron hits the superconductor then the 
retro-reflected hole diffracts on the scatterer and makes a loop with 
repetition number $r$ as described before Eq.\ (\ref{lr}). Next, 
the hole diffracts off the scatterer then hits the superconductor on
which it converts back to an electron. Finally, the electron goes back 
to the entrance lead.

The above results represent the first theoretical study of the path-length
spectra of a normal superconductor
mesoscopic system with diffractive scattering. 
We have demonstrated that the PLS of such systems differs fundamentally
from that of
normal systems, due to the appearance of peaks at negative lengths.
To highlight this effect, we have analyzed the simplest possible diffractive
scatterer. 
Since the appearance of negative lengths in PLS is the {\em direct} 
consequence of the presence of diffractive scatterers it is desirable 
to study other types of diffractive centers such as corners.
There is a growing interest in studying the role of 
diffractive scatterers in normal mesoscopic systems\cite{Jalabert}, 
therefore the extension  
to normal superconductor systems may become a new playground both 
in the semiclassical 
theory of scattering processes and level statistics of these systems.
For the future it will also be of interest to examine the amplitudes of
the negative-length peaks in more complex geometries, such as those of
\cite{Colin1},
to examine the effect of a tunnel junction at the NS interface and the role of
order parameter symmetry.

This work was supported by the EU. TMR within the programme 
``Dynamics of Nanostructures'' jointly with OMFB, the Hungarian 
Science Foundation OTKA  T025866, 
the Hungarian Ministry of Education (FKFP 0159/1997).

\begin{figure}
{\centerline{\leavevmode \epsfxsize=8.5cm \epsffile{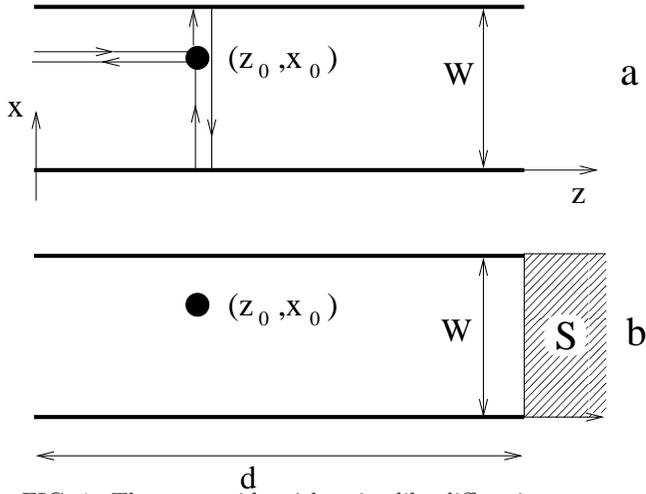 }}}
\caption{The waveguide with point-like diffractive scatterer (a) opened 
and (b) a superconductor attached to the exit lead 
(left side of the waveguide). 
The origin of the coordinate system $(z,x)$ is at the left bottom corner 
of the waveguide. In Fig.\ 1a  one possible trajectory is shown. For clarity 
the forward and backward paths are shifted.  
\label{abra1}}
\end{figure} 

\begin{figure}
{\centerline{\leavevmode \epsfxsize=8.5cm \epsffile{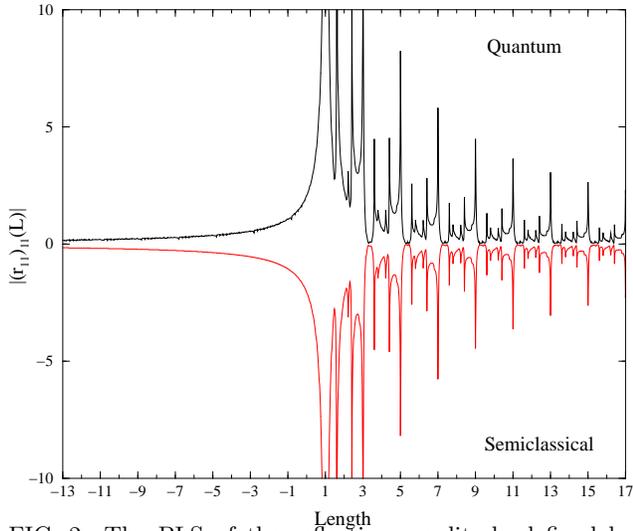 }}}
\caption{The PLS of the reflection amplitude defined by 
Eq.\ (\ref{PLS}) is obtained from quantum mechanical and 
semiclassical calculations. 
The width of the waveguide, $W=1.0.$ The strength of the Dirac delta 
potential, $\lambda=10.0$, and its position is located at 
$(z_0,x_0)=(0.5,0.7)$. 
\label{Q-szemi-N}}
\end{figure} 

\begin{figure}
{\centerline{\leavevmode \epsfxsize=8.5cm \epsffile{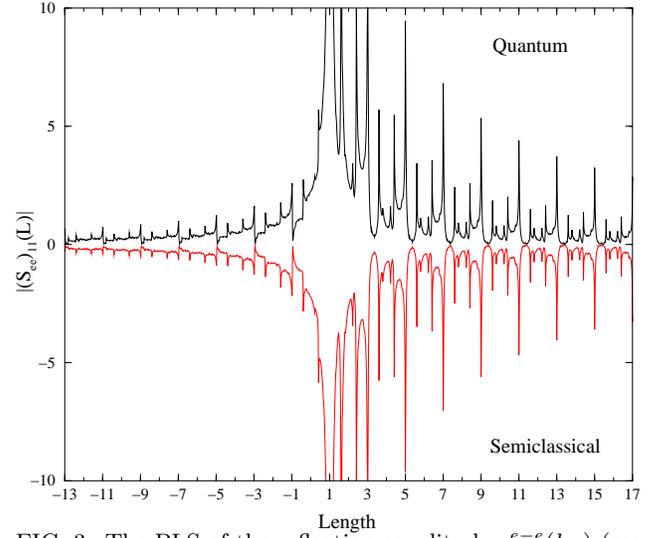 }}}
\caption{The PLS of the reflection amplitude  $s_{11}^{e-e}(k_F)$ (see 
Eq.\ (\ref{hetes})) is obtained from quantum and 
semiclassical calculations. For parameters see Fig.\ 2. 
\label{Q-szemi-S}}
\end{figure}

\begin{figure}
{\centerline{\leavevmode \epsfxsize=8.5cm \epsffile{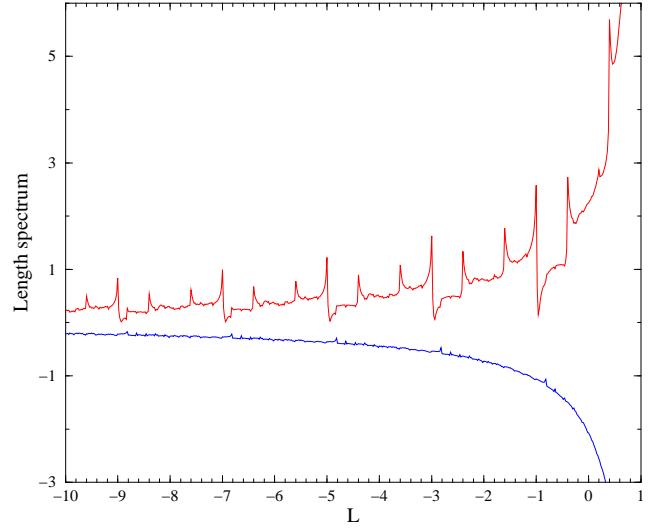 }}}
\caption{The PLS of the reflection amplitude $r_{11}(k_F)$ of 
the normal system (lower part) and $s_{11}^{e-e}(k_F)$
given in Eq.\ (\ref{hetes}) (upper part) are plotted. 
The most pronounced peaks with negative lengths 
come from the family $L= 2z_0 -l_r$, where $l_r$ is 
given in Eq.\ (\ref{lr}), namely 
$L=-0.4,-1.0,-1.6,-2.4,-3.0,-3.6,-4.4,-5,\cdots$ .
For parameters see Fig.\ 2. 
\label{Q-NS}}
\end{figure} 


\begin{thebibliography}{99}
\bibitem{6} C. M. Marcus, A. J. Rimberg, R. M. Westervelt, P. F.
Hopkins and A. C. Gossard, Phys. Rev. Lett. {\bf 69}, 506 (1992);
C. M. Marcus, R. M. Westervelt, P. F. Hopkins and A. C. Gossard, Chaos
{\bf 3}, 643 (1993).
\bibitem{7} M. W. Keller, O. Millo, A. Mittal, D. E. Prober and R. N.
Sacks, Surf. Sci. {\bf 305}, 501 (1994).
\bibitem{4} K. B. Efetov, Adv. Phys. {\bf 32}, 53 (1983).
\bibitem{5} R. Bl\"umel and U. Smilansky, Phys. Rev. Lett. {\bf 64},
241 (1990).
\bibitem{Carlo1}C.~W.~J. Beenakker, Rev. Mod. Phys. {\bf 69},  731  (1997).
\bibitem{B1} R. A. Jalabert, H. U. Baranger, and A. D. Stone,
Phys. Rev. Lett. {\bf 65}, 2442 (1990).
\bibitem{B2} H. U. Baranger, R. A. Jalabert, and A. D. Stone,
Chaos {\bf 3}, 665 (1993). 
\bibitem{Burgerdorf1}
H. Ishio and J. Burgd{\"o}rfer, Phys. Rev. B {\bf 51},  2013  (1995).
\bibitem{Burgerdorf2}
L. Wirtz, J. Tang, and J. Burgd{\"o}rfer, Phys. Rev. B {\bf 56},  7589  (1997).
\bibitem{Burgerdorf3}
J. Tang, L. Wirtz, and J. Burgd{\"o}rfer, Phys. Rev. B {\bf 57},  9875  (1998).
\bibitem{Delos}
C.~D. Schwieters, J.~A. Alford, and J.~B. Delos, Phys. Rev. B {\bf 54},  10652
  (1996).
\bibitem{Lin-semi}
W.~A. Lin and R.~V. Jensen, Phys. Rev. B {\bf 53},  3638  (1996).
\bibitem{Andreev}
A.~F. Andreev, Zh. Eksp. Teor. Fiz. {\bf 46},  1823  (1964), [Sov. Phys. JETP,
  {\bf 19}, 1228 (1964)].
\bibitem{Carlo-konyv}
C.~W.~J. Beenakker,  in {\em Mesoscopic Physics, Les Houches Summer School},
  edited by E. Akkermans, G. Montambaux, J.~L. Pichard, and J. Zinn-Justin
  (Elsevier Science B. V., Amsterdam, 1995).
\bibitem{Colin1}
C.~J. Lambert and R. Raimondi, J. Phys. Condens. Matter {\bf 10},  901  (1998).
\bibitem{Chaos-cikk}G. Vattay, J. Cserti, G. Palla, and G. Sz\'alka, Chaos, Solitons \& Fractals
  {\bf 8},  1031  (1997); (cond-mat/9703088).
\bibitem{Keller}  J. B. Keller, J. Opt. Soc. Am. {\bf 52}, 116 (1962).
\bibitem{Diffract} G. Vattay, A. Wirzba and P. E. Rosenqvist, Phys. Rev.
Lett. {\bf 73}, 2304 (1994); N. Pavloff and C. Schmit, Phys. Rev. Lett.
{\bf 75}, 61 (1995) (Erratum ibid. p.3779).; H. Bruus and N. D. Whelan,
Nonlinearity {\bf 9}, 1023 (1996). 
\bibitem{Exner} P. E. Rosenqvist, N. D. Whelan and A. Wirzba,
J. Phys. A {\bf 29}, 5441 (1996); 
P. Exner and P. \'Seba, Phys. Lett. A {\bf 222}, 11
(1996); P. Dahlqvist and G. Vattay, J. Phys. A {\bf 31}, 6333 (1998).
\bibitem{BB} See \cite{Carlo1} p. 789.
\bibitem{Crossover}
See Eqs.\ (7) and (6) in
J. Cserti, G. Sz\'alka, and G. Vattay, Phys. Rev. B {\bf 57},  R15092
(1998).
\bibitem{Jalabert} R. A. Jalabert, cond-mat/9912038; 
J. S. Hersch, M. R. Haggerty, E. J. Heller, nlin.CD/0003023; 
J. S. Hersch, M. R. Haggerty, and E. J. Heller, Phys. Rev. Lett., 
83, 5342 (1999); M. Sieber, nlin.CD/0003019; 
M. Sieber, J.\ Phys. A {\bf 32}, 7679 (1999); 
E. Bogomolny, P. Leboeuf, C. Schmit,  nlin.CD/0003016.
\end{thebibliography}
\end{document}